\documentstyle[12pt]{article}
\newskip\humongous \humongous=0pt plus 1000pt minus 1000pt

\newif\ifdtup

\def\pr#1{#1^\prime}

\def\beq{\begin{equation}}
\def\eeq{\end{equation}}
\def\eq{\beq\eeq}
\def\beqn{\begin{eqnarray}}
\def\eeqn{\end{eqnarray}}
\relax

\def\dotx{\dotx{\dot\overline{x}}}

\relax

\catcode`\@=11

\@addtoreset{equation}{section}
\def\theequation{\thesection\arabic{equation}}

\def\@normalsize{\@setsize\normalsize{15pt}\xiipt\@xiipt
\abovedisplayskip 14pt plus3pt minus3pt%
\belowdisplayskip \abovedisplayskip
\abovedisplayshortskip \z@ plus3pt%
\belowdisplayshortskip 7pt plus3.5pt minus0pt}

\def\small{\@setsize\small{13.6pt}\xipt\@xipt
\abovedisplayskip 13pt plus3pt minus3pt%
\belowdisplayskip \abovedisplayskip
\abovedisplayshortskip \z@ plus3pt%
\belowdisplayshortskip 7pt plus3.5pt minus0pt
\def\@listi{\parsep 4.5pt plus 2pt minus 1pt
     \itemsep \parsep
     \topsep 9pt plus 3pt minus 3pt}}

\@twosidetrue

\relax

\catcode`@=12

\catcode`\@=11

\def\section{\@startsection{section}{1}{\z@}{3.5ex plus 1ex minus
   .2ex}{2.3ex plus .2ex}{\large\bf}}

\def\thesection{\arabic{section}.}

\def\appendix{\setcounter{section}{0}
 \def\thesection{APPENDIX \Alph{section}:}
 \def\theequation{\Alph{section}.\arabic{equation}}}

\def\ps@headings{\def\@oddfoot{}\def\@evenfoot{}
\def\@oddhead{\hbox{}\hfill
 \makebox[.5\textwidth]{\raggedright\ignorespaces --\thepage{}--
 \hfill {}}}  
\def\@evenhead{\@oddhead}
\def\subsectionmark##1{\markboth{##1}{}}
}

\ps@headings

\catcode`\@=12

\def\figcap{\section*{Figure Captions\markboth
 {FIGURECAPTIONS}{FIGURECAPTIONS}}\list
 {Fig. \arabic{enumi}:\hfill}{\settowidth\labelwidth{Fig. 999:}
 \leftmargin\labelwidth
 \advance\leftmargin\labelsep\usecounter{enumi}}}
 \relax
\def\tablecap{\section*{Table Captions\markboth
 {TABLECAPTIONS}{TABLECAPTIONS}}\list
 {Table \arabic{enumi}:\hfill}{\settowidth\labelwidth{Table 999:}
 \leftmargin\labelwidth
 \advance\leftmargin\labelsep\usecounter{enumi}}}
 \relax
\def\reflist{\section*{References\markboth
 {REFLIST}{REFLIST}}\list
 {[\arabic{enumi}]\hfill}{\settowidth\labelwidth{[999]}
 \leftmargin\labelwidth
 \advance\leftmargin\labelsep\usecounter{enumi}}}
 \relax

\catcode`\@=11

\def\ps@headings{\def\@oddfoot{}\def\@evenfoot{}
\def\@oddhead{\hbox{}\hfill
 \makebox[.5\textwidth]{\raggedright\ignorespaces --\thepage{}--
 \hfill {}}}    
\def\@evenhead{\@oddhead}
\def\subsectionmark##1{\markboth{##1}{}}
}

\ps@headings

\relax

\relax
\def\pl#1#2#3{{\it Phys. Lett. }{\bf #1}(19#2)#3}
\def\zp#1#2#3{{\it Z. Phys. }{\bf #1}(19#2)#3}
\def\prl#1#2#3{{\it Phys. Rev. Lett. }{\bf #1}(19#2)#3}

\def\pr#1#2#3{{\it Phys. Rev. }{\bf #1}(19#2)#3}
\def\np#1#2#3{{\it Nucl. Phys. }{\bf #1}(19#2)#3}

\relax

\begin{document}
\def\theequation{\arabic{equation}}
\newcommand\as{\alpha_S}
\newcommand\aem{\alpha_{\rm em}}
\newcommand\refq[1]{$^{[#1]}$}
\newcommand\avr[1]{\left\langle #1 \right\rangle}
\newcommand\lambdamsb{
\Lambda_5^{\rm \scriptscriptstyle \overline{MS}}
}
\newcommand\qqb{{q\overline{q}}}
\newcommand\asb{\as^{(b)}}
\newcommand\qb{\overline{q}}
\newcommand\sigqq{\sigma_{q\overline{q}}}
\newcommand\siggg{\sigma_{gg}}
\newcommand\MSB{{\rm \overline{MS}}}
\def \eq {e_{\scriptscriptstyle Q}}
\def \muf {\mu_{\scriptscriptstyle F}}
\def \mur {\mu_{\scriptscriptstyle R}}
\def \muo {\mbox{$\mu_0$}}
\def \pt  {\mbox{$p_{\rm T}$}}
\def \ptg {\mbox{$p_{\rm T}^{q\overline{q}}$}}
\def \xf  {\mbox{$x_{\rm F}$}}
\def \dphi{\mbox{$\Delta\phi$}}
\def \dy  {\mbox{$\Delta y$}}
\def \pim {\mbox{$\pi^-$}}
\def \epem {\mbox{$e^+e^-$}}
\def \mc   {\mbox{$m_c$}}
\def \mb   {\mbox{$m_b$}}
\def \mqq   {\mbox{$M_{q\overline{q}}$}}
\def \tot   {{\rm tot}}
\newcommand\qq{{\scriptscriptstyle Q\overline{Q}}}
\newcommand\cm{{\scriptscriptstyle CM}}
\setcounter{topnumber}{10}
\setcounter{bottomnumber}{10}
\renewcommand\topfraction{0}
\renewcommand\textfraction{0}
\renewcommand\bottomfraction{1}
\begin{titlepage}
\nopagebreak
\vspace*{-1in}
{\leftskip 11cm
\normalsize
\noindent
\newline
CERN-TH.6864/93\\
GeF-TH-12/93

}
\vfill
\begin{center}
{\large \bf \sc On the Determination of the Gluon Density of the Proton \\
 from Heavy-Flavour Production at HERA}
\vfill
{\bf Stefano Frixione}
\vskip .3cm
{Dip. di Fisica, Universit\`a di Genova, and INFN, Sezione di Genova,
Genoa, Italy}\\
\vskip .6cm
{\bf Michelangelo L. Mangano}
\vskip .3cm
{INFN, Scuola Normale Superiore and Dipartimento di Fisica, Pisa, Italy}\\
\vskip .6cm
{\bf Paolo Nason\footnotemark}
\footnotetext{On leave of absence from INFN, Sezione di Milano, Milan, Italy.}
and
{\bf Giovanni Ridolfi\footnotemark}
\footnotetext{On leave of absence from INFN, Sezione di Genova, Genoa, Italy.}
\vskip .3cm
{CERN TH-Division, CH-1211 Geneva 23, Switzerland}
\end{center}
\vfill
\nopagebreak
\begin{abstract}
{\small
Using a recent next-to-leading-order calculation of the
photoproduction double differential cross section for
heavy quarks, we study the possibility of extracting the gluon density
of the proton from heavy-quark photoproduction data. We discuss the
theoretical uncertainties connected with this method, and we
conclude that they are well under control in a wide
$x$ domain.}
\end{abstract}
\vfill
CERN-TH.6864/93
\newline
April 1993    \hfill
\end{titlepage}
The gluon density of the proton is usually extracted from fits to deep
inelastic scattering data.
There it affects the evolution of the quark density, according
to the Altarelli-Parisi\refq{\ref{AltarelliParisi}}
evolution equation. Its determination is indirect, and
to some extent, it is influenced by the particular parametrization chosen for
the fits. A direct determination is therefore highly desirable. In
hadron-hadron collisions, processes such as direct
photon\refq{\ref{DirectPhoton1},\ref{DirectPhoton2}}
and jet\refq{\ref{Jet}} production are
directly sensitive to the gluon distribution. Although these processes provide
further constraints on the gluon density, they do not allow a precision
comparable with the DIS experiments.

The $ep$ collider HERA offers new opportunities to measure the gluon density
directly. Since the initial state is simpler than in hadron-hadron collisions,
the gluon density enters in a simpler way. Furthermore, various
methods can be used, thus allowing for consistency checks. Some of the
methods
are reviewed in ref.~[\ref{HeraWorkshop}]. The most promising one is based
upon the measurement of the longitudinal structure
function\refq{\ref{FLmethod}}. It will probably allow a determination of the
gluon density in the $x$ range between $10^{-3}$ and $5\times 10^{-2}$ for
$25$ GeV$^2$ $<Q^2<150$ GeV$^2$.

There are a number of methods that rely upon the study of
hard production processes.
At HERA, these processes arise both from the pure photon
component and from the hadronic (sometimes called resolved) component
of the photon. In jet production, the effect of the
latter is strongly enhanced, because the
typical elementary process $gg\to gg$ is about 20 times larger than
the processes $\gamma g\to q\bar{q}$ and $\gamma q\to q g$.
One possible strategy to
suppress the hadronic component is to consider only off-shell photon
jet production. This suppresses the rate, but eliminates the hadronic
component. One should then disentangle the two pure photon processes
$\gamma g\to q\bar{q}$ and $\gamma q\to q g$.
The other possibility is to require heavy-quark production. In the
hadronic component only the two subprocesses
(at leading order) $gg \to Q\bar{Q}$ and
$q\bar{q}\to Q\bar{Q}$ contribute, and they are
two orders of magnitude smaller\refq{\ref{KunsztP}}
than the typical process $gg\to gg$. The hadronic component is not
eliminated completely with this method, but there are large kinematical
ranges in which it is negligible.

Along this line a popular method is to consider $J/\psi$
production\refq{\ref{JPSImethod}}. Although experimentally very convenient,
this method suffers from several theoretical ambiguities, essentially due to
difficulties in the computation of the $J/\psi$ production cross section.
Open-charm production is a more promising method\refq{\ref{OpenCmethod}}.
It is less
affected by theoretical ambiguities, and it is sensitive to the
gluon density in the small-$x$ region. In ref.~[\ref{OpenCmethod}]
a study of the feasibility of measuring the gluon density using open charm
production was performed, using leading-order cross sections. It was concluded
that $f^{(p)}_g(x)$ can be measured for $x$ in the range between $10^{-1}$ and
$10^{-3}$.

In order to extract the gluon density from open-charm production to
next-to-leading accuracy, a next-to-leading calculation of the
heavy-quark photoproduction cross section is needed.
Such calculations for single-inclusive
cross sections (i.e. integrated over the momentum of one of the heavy quarks)
have been available for some time\refq{\ref{EllisNason},\ref{SmithNeerven}}.
They are not, however, very useful for our purpose, since they do not allow
for the full reconstruction of the $x$ variable of the parton in the proton
from the final-state observables.

We have recently completed a next-to-leading calculation of the double
differential heavy-quark photoproduction cross section.  This calculation,
which will be described in more detail in a forthcoming
publication\refq{\ref{FMNR}}, is an extension of the results presented in
refs.~[\ref{MNR}] and [\ref{MNRFT}]. In the present work, we will apply this
result to the problem of extracting the gluon density of the proton from
open-charm production. The main goal of this
work is to show that the theoretical
uncertainties associated with this procedure are well under control, and to
assess their magnitude. The possible sources of uncertainty in the problem are
the following: radiative corrections of even higher order (i.e. ${\cal
O}(\aem\as^3)$), not included in our calculation, the poor knowledge of the
value of the heavy-quark mass, uncertainties on the value of $\Lambda_{\rm
QCD}$, and the presence of a hadronic component of the photon.
Furthermore, one should not forget the large hadronization effects
that are usually found in
charm production. There are good reasons to believe that these effects are
smaller in the photoproduction than in the hadroproduction case. We will also
argue that there are several ways of getting rid of the major hadronization
ambiguities.

We begin by writing the heavy-quark cross section at the leading order
in the following form
\beq
\frac{d \sigma^{(0)}}{d y_{\qq} \, dM_{\qq}^2 }=
x_g\frac{d \sigma^{(0)}}{d x_g \, dM_{\qq}^2 }= \frac{1}{E^2}\;
f^{(e)}_{\gamma}(x_\gamma,\mu_0^2)
f^{(p)}_{g}(x_g,\muf^2)
\hat{\sigma}^{(0)}_{\gamma g}(M_{\qq}^2), \label{SigmaBorn}
\eeq
where $M_{\qq}$ is the invariant mass of the heavy-quark pair, and
$y_{\qq}$ is the rapidity of the pair in the electron-proton
centre-of-mass frame (we choose positive $y$ in the direction of
the incoming photon).
$E$ is the electron-proton centre-of-mass energy, and
\beqn
x_\gamma&=&\frac{M_{\qq} }{E} \exp(y_{\qq}) \\
x_g&=&\frac{M_{\qq} }{E} \exp(-y_{\qq}). \label{xgdef}
\eeqn
The function $f^{(e)}_{\gamma}$ is the photon density in the electron in the
Weizs\"acker-Williams approximation,
\beq \label{WW}
f^{(e)}_{\gamma}(x,\mu_0^2)=\frac{\aem}{2\pi}\frac{1+(1-x)^2}{x}
\log\frac{\mu_0^2 (1-x)}{m_e^2 x^2}
\eeq
and $f_g^{(p)}$ is the density of gluons in the proton.
The Born level partonic cross section
$\hat{\sigma}_{\gamma g}^{(0)}(s)$ is given by
\beq \label{BornParton}
\hat{\sigma}_{\gamma g}^{(0)}(s)=\frac{\eq^2\aem\as(\mur)}{m^2}
\frac{\pi\beta\rho}{4}\left[\frac{3-\beta^4}{\beta}\log\frac{1+\beta}{1-\beta}
-4+2\beta^2 \right],
\eeq
where $m$ is the heavy-quark mass, $\eq$ is its charge in electron
charge units, and
\beq
\rho=\frac{4m^2}{s},\quad \beta=\sqrt{1-\rho} \; .
\eeq
The renormalization scale is set to
$\mur=\mu_0$, and the reference scale $\mu_0$ is defined by
\beq
\mu_0=\sqrt{(p_T^2+\bar{p}_T^2)/2+m^2},
\eeq
where $p_T$ and $\bar{p}_T$ are the transverse momenta of the heavy
quark and of the heavy antiquark.
The factorization scale for the proton is set to $2\mu_0$.

Assuming that we identify the left-hand side of eq.~(\ref{SigmaBorn}) with the
data, we can invert the equation, and get a first determination of $f^{(p)}_g$:
\beq
f^{(0)}_g(x_g,\muf^2)=
x_g\frac{d \sigma^{\rm data}}{d x_g \, dM_{Q\bar {Q}}^2 } \frac{E^2}
{f^{(e)}_{\gamma}(x_\gamma,\mu_0^2)
\hat{\sigma}^{(0)}_{\gamma g}(M_{\qq}^2) }.
\eeq
The inclusion of radiative corrections does not pose any problem.
We write the full cross section as
\beq \label{SigmaNL}
x_g \frac{d \sigma}{d x_g \, dM_{\qq}^2 }=
\frac{1}{E^2}\;f^{(e)}_{\gamma}(x_\gamma,\mu_0^2)
f^{(p)}_{g}(x_g,\muf^2)
\hat{\sigma}^{(0)}_{\gamma g}(M_{\qq}^2)
+\Delta(f^{(p)}_g,x_g,M_{\qq}^2),
\eeq
where $\Delta$ represents all the radiative effects. In $\Delta$ we have also
indicated explicitly the functional dependence upon the gluon density in the
proton. The light quarks, which enter at the next-to-leading order via the
$\gamma q\to q Q\bar{Q}$ process, give a small contribution (less than 5\% for
all values of $x_g$ and $M_\qq$ considered here). The $\Delta$ term is given by
\beqn
\Delta(f^{(p)}_g,x_g,M_{\qq}^2)
&=&\sum_j \int dx\,dx_\gamma
 f^{(e)}_{\gamma}(x_\gamma,\mu_0^2)
f^{(p)}_j(x,\muf^2)
\aem\as^2(\mur)
\nonumber \\
&& \;\;\;\; \times
\frac{d\hat\sigma^{(1)}_{\gamma j} }
{d M_{\qq}^2 d \hat{y}_{\qq}}
(x_\gamma x E^2,M_{\qq}^2,\hat{y}_{\qq},\muf,\mur,\mu_\gamma),
\eeqn
where
\beq
\hat{y}_{\qq}=\frac{1}{2}\log\frac{M_{\qq}^2}
{x_\gamma x_g E^2}
\eeq
is the heavy-quark pair rapidity in the partonic centre of mass frame. The
factorization scale of the photon $\mu_\gamma$ is set to 1~GeV (we
verified that the results are rather insensitive to the choice of
$\mu_\gamma$).
For a complete
discussion of all the scale choices we refer the reader to a forthcoming
publication\refq{\ref{FMNR}}.

We now write the full $f^{(p)}_g$ as
\beq
f^{(p)}_g(x,\mu^2)=f^{(0)}_g(x,\mu^2)+f^{(1)}_g(x,\mu^2),
\eeq
where the second term is the next-to-leading correction, and plug it back into
eq.~(\ref{SigmaNL}).
We get
\beq
f_g^{(1)}(x_g,\muf^2)=-\frac{E^2\Delta(f^{(0)}_g,x_g,M_{\qq}^2)}
{f^{(e)}_\gamma(x_\gamma,\mu_0^2)
\;\hat{\sigma}^{(0)}_{\gamma g}(M_{\qq}^2) }.
\label{f1correction}
\eeq
We have neglected
the $f_g^{(1)}$ piece contained in the $\Delta$ term,
the corresponding contribution being of order $\aem\as^3$.
It could also be easily incorporated by iterating
eq.~(\ref{f1correction}), using the full structure function in the
right-hand side.
In the simple illustration we have given, we have
integrated the cross section over all variables but $x_g$ and
$M_\qq$. In general, in realistic experimental configurations,
other cuts will be applied to the
data. The procedure for the extraction of $f^{(p)}_g$ outlined above
can be carried out also in this
case, provided the same cuts as applied to the hadronic final states in
the data are also applied to the partonic final state in the
calculation. One interesting possibility is to apply this procedure
to the invariant mass of the two-jet system, instead of the invariant
mass of the heavy-quark pair. This makes no difference at the Born
level, but next-to-leading corrections will turn out to be different,
and one should compute them with an appropriate jet definition (which
should match the jet definition used in the analysis of the data) in
order to get a meaningful answer (see refs.~[\ref{jetdefinitions}]
for a discussion of this point). Furthermore, with this procedure
there would be no need to identify both charmed mesons, and the result
would be less sensitive to fragmentation effects.

We now begin with a study of the charm differential distribution in the
variable $x_g$ (as defined in eq.~(\ref{xgdef})) at HERA,
for various cuts in $M_\qq$. These are the distributions
which are relevant to the extraction of the gluon density at HERA.
All our results are obtained for $\sqrt{S}=$314~GeV.
In fig.~\ref{FullBornFig} we plot the leading-order and next-to-leading-order
pure photon cross
section (throughout this work, we use the proton structure function set HMRS~B
of ref.~[\ref{HMRSB}]).
{}From the figure we first notice
that radiative corrections are moderate, but not negligible. This is an
indication that the perturbative expansion is reliable in this case.
We also observe that the effect of the radiative corrections cannot be
described by a simple $K$-factor.
{}From fig.~\ref{FullBornFig}, neglecting for the moment the contribution of
the
hadronic component of the photon, we can infer that the distribution is
sensitive to the
gluon density of the proton down to $x_g$ of the order of a few times
$10^{-4}$.

We now examine the effect of the hadronic component of the photon.
This contributes another term to the cross section, given by
\beq
d \sigma_{\scriptscriptstyle H}
=
\sum_{ij} \int dx\,dx_\gamma \,\left[
\int^1_{x_\gamma} f^{(e)}_{\gamma}(z,\mu_0^2)
\; f^{(\gamma)}_i\left(\frac{x_\gamma}{z},\mu_\gamma^2\right)\frac{dz}{z}
\right]
f^{(p)}_{j}(x,\muf^2) \;
d \hat{\sigma}_{ij} \label{SigmaH}.
\eeq
This is formally of the same order as eq.~(\ref{SigmaBorn}), since the photon
parton densities $f^{(\gamma)}_i$ are of order $\aem/\as$. The contribution of
the hadronic component, including next-to-leading orders (computed using the
program of ref.~[\ref{MNR}]) is displayed in fig.~\ref{FullHadrFig} for two
different parametrizations of the photon structure function. We have chosen
the set LAC~1 (ref.~[\ref{Abramowicz}]) and the set ACFGP~HO-mc
(ref.~[\ref{Aurenche}]). These two sets have extreme gluon distributions  (as
can be easily seen from fig.~10 of ref.~[\ref{Besch}]), and they may
therefore give an idea of the magnitude of the hadronic component contribution
(experimental results from HERA will help to better
specify the value of this component).
We see from fig.~\ref{FullHadrFig} that the hadronic component gives a large
contribution for the smallest invariant-mass cut. Even in this case,
however, there is a region of small $x_g$ in which the hadronic component
is negligible with respect to the pure photon term.
When we increase the invariant mass cut, we notice that the hadronic
component decreases faster than the pure photon contribution.
This is due to the fact that in the hadronic component,
when we increase the invariant mass, the production process
is suppressed by the small value of the gluon density of the photon at large
$x$.
We therefore see that a large region in $x_g$ is accessible by this
method. By pushing the invariant-mass cut to 20~GeV we can reach the
region of $x_g=10^{-1}$. Observe that statistics will not be a problem
even at these large invariant-mass cuts. With 100~pb$^{-1}$ of
integrated luminosity, there would be about $10^5$ events
per bin in the case of a 20~GeV cut (before accounting for experimental
efficiencies).
We therefore find that the conclusions of ref.~[\ref{OpenCmethod}]
are not spoiled
by theoretical ambiguities due to higher-order effects or to the
hadronic component of the photon.
We then conclude that the
theoretical uncertainties in the heavy-flavour cross section,
in the range of $10^{-3}<x_g<10^{-1}$, are small
enough to allow for a determination of the gluon density in the proton.

The authors of ref.~[\ref{SmithNeervenRiemersma}] reach a conclusion
that contrasts with ours, that is to say that
the ambiguities related to the hadronic component of the photon
spoil the predictivity of the method. This conclusion is based upon a
single inclusive calculation.
They find that in particular the set LAC~3 of the photon structure
functions of
ref.~[\ref{Abramowicz}] gives a hadronic contribution that competes with
the pure photon contribution for all rapidities.
We find that even with the LAC~3 set, when looking at the double
differential quantities we have considered, it is still possible to
perform the extraction of the gluon density in the proton, although in
a more restricted $x_g$ range. However, we have chosen not to include
this structure function set in our analysis for the following reasons.
The LAC~3 set has an
unphysically hard gluon structure function.
For $Q^2=5\,\mbox{GeV}^2$, it peaks at $x=0.9$
and carries 70\% of the total momentum of
the hadronic component of the photon. Because of this very pronounced
peak, it is roughly as hard as the pure photon.
We have examined the origin of this behaviour.
As can be seen from ref.~[\ref{Abramowicz}], the set LAC~3, unlike
the sets LAC~1 and LAC~2, is obtained by fitting data for $F_2^\gamma(x,Q^2)$
at $Q^2$ values down to $1.31\,\mbox{GeV}^2$ (for the other sets the fits
start at $Q^2=4.3\,\mbox{GeV}^2$). From $Q^2=1.3\,\mbox{GeV}^2$ to
$Q^2=4.3\,\mbox{GeV}^2$, $F_2^\gamma(x,Q^2)$ grows rapidly in the
region $x>0.2$. In order to reproduce this growth
using the QCD evolution, one is forced to assume a very hard gluon
component, which under evolution will feed down a large quark component
at moderate values of $x$. On the other hand, at this low $Q^2$
values, there may be other
reasons, having nothing to do with the QCD evolution, that can cause
the observed growth. For example,
new thresholds for single resonances or resonance-pair production can
be opened.
After all, at $Q^2=1.3$ GeV$^2$ and $x=0.3$ the mass of
the produced hadronic system
is only of 1.74~GeV. We therefore believe that these data are more
consistent with a change in regime for $F_2^\gamma$, from a low-energy
VMD behaviour to a high-energy point-like behaviour, than with
perturbative evolution; one should thus
not attempt to fit it with QCD evolution alone.
Indications that one should use $Q^2$ values
larger than 4~GeV$^2$ were also given in
ref.~[\ref{Vieira}]. Furthermore, the authors
of ref.~[\ref{Abramowicz}] also express some scepticism with respect
to their set 3.
Finally, experimental results on jet production in photon-photon
collisions\refq{\ref{LAC3failsTOPAZ}} and photon-hadron
collisions\refq{\ref{LAC3failsH1}} clearly disfavour the set LAC~3.

We will now turn to a discussion of the other theoretical
uncertainties involved in our procedure, in order to estimate the
precision of the method. One source of uncertainty is due to the
truncation of the perturbative expansion. We estimate this
uncertainty by varying the renormalization and
factorization scales by a factor of two above and below their
reference value. The result is plotted in fig.~\ref{ScaledepFig}.
We see that the effect of the variation of the factorization scale
is moderate, while the renormalization scale uncertainty amounts to
an uncertainty of $\pm$10 to 20\% on the result.

In fig.~\ref{MassdepFig} we show  the dependence of our result upon the heavy
quark mass. Even in this case the variation of the cross section is of
the order of 10 to 20\%. We should remind, however, that the mass of
the heavy quark (unlike the renormalization scale) is a physical
parameter that enters the cross section. It therefore makes sense
to reduce this uncertainty by using values of the mass that
give a good fit of the data.
There are of course also uncertainties associated with the error
on the present determination of $\Lambda_{QCD}$.
In practice, it will be more convenient to measure $\as f^{(p)}_g(x)$,
which is much less sensitive to the value of $\Lambda_{QCD}$.

We observe that perturbation theory alone makes a prediction for the
ratio of our differential distributions for different mass cuts in the
$x_g$ region in which the hadronic component is small.
In leading order and up to scaling violation, the gluon density in the proton
cancels out in this ratio.
We present a plot of these ratios in fig.~\ref{RatioFig}.
As can be seen from the figure, the mass and scale dependence of the
ratios for the pure photon contribution is negligible, while
significant changes can be observed when the hadronic component
is included. This in principle might allow for an independent
check of the hadronic structure of the photon.

As suggested in ref.~[\ref{Halzen}], an additional help to separate
the pure from the hadronic photon contributions comes from tagging
the photon energy by measuring the energy of the recoiling electron
in the small-angle luminosity monitors. This approach might have, in general,
problems at next-to-leading order: processes such as
$\gamma q \to q Q\bar{Q}$, the light quark being emitted
preferentially in the direction of the photon, will look and will be
reconstructed as hadronic photon events. As was mentioned previously, however,
the contribution of these processes is numerically negligible.
The photon-tagging technique could then help to further constrain the gluon
density of the photon.

We conclude by giving in table~1 the total cross sections for $b$ and $c$
production at HERA. Observe that
the sensitivity to the variation of the
scale and mass parameters is much smaller for $b$ than for $c$
production. This suggests that by using $b$ production a
better precision may be achieved
in the determination of the gluon density. On the
other hand the $x_g$ region covered is smaller. We see in fact that
using the structure function set MSRD-- (a set with a gluon density
more singular at small $x$) the charm cross section grows
by 80\%, while the $b$ cross section grows by less than 10\% with
respect to our central value.

Observe the large difference in the hadronic component when using the
two different sets of photon structure functions.
When going from the ACFGP to the LAC~1 set, the hadronic component of
the charm cross section grows by a factor of 9, while for bottom
it grows only by a factor of 3. This difference is mainly due to
the small $x$ growth of the gluon density in the photon. It is likely
that the charm cross section and distributions at HERA will also help
to constrain the gluon density in the photon at small $x$.

\begin{table}
\begin{center}
\begin{tabular}{|l||c|c|} \hline
& $\sigma_{c\bar{c}}$ ($nb$) & $\sigma_{b\bar{b}}$ ($nb$) \\
\hline\hline
 Central value        &423  & 4.37  \\ \hline
 $\mur=2\mu_0$        &362  & 3.90  \\ \hline
 $\mur=\mu_0/2$       &489  & 4.94  \\ \hline
 $\muf=2\mu_0 $       &  -  & 4.50  \\ \hline
 $\muf=\mu_0/2$       &  -  & 4.13  \\ \hline
 Low mass             &705  & 5.16  \\ \hline
 High mass            &270  & 3.72  \\ \hline
 $\Lambda_4=100\;$MeV &337  & 3.66  \\ \hline
 $\Lambda_4=300\;$MeV &453  & 4.66  \\ \hline
 MRSD--                &765  & 4.65  \\ \hline
 Hadr. comp. LAC 1    &723  & 3.10  \\ \hline
 Hadr. comp. ACFGP    &80.5 & 0.91  \\ \hline
\end{tabular}
\end{center}
\caption[]{\label{SigmaHera}
Charm and bottom pair-production cross sections at HERA.
The pure photon component and the hadronic component are shown
separately. The central values for the scales are $\mur=m_c$ for
charm and $m_b$ for bottom, $\muf=2m_c$ for charm and $m_b$ for
bottom. The central mass value for charm used here is 1.5~GeV,
the low value is 1.2~GeV and  the high value is 1.8 GeV. For bottom we set
$m_b=4.75\;$GeV, the low value is 4.5~GeV and the high value is 5~GeV.
The structure function set for the central value is HMRSB, with
$\Lambda_4=190\;$MeV. When studying the sensitivity to $\Lambda$ we
use two HMRSB fits\refq{\ref{PDFLIB}} obtained with the
values of $\Lambda$ indicated in the table.
We also show the values obtained using the MRSD-- structure functions,
which have stronger enhancement at small $x$.
For the hadronic component only the central values are used, since
the uncertainty due to the choice of the photon structure function
is by far the largest.}
\end{table}

\clearpage
\begin{reflist}
\item\label{AltarelliParisi}
   G.~Altarelli and G.~Parisi, \np{B126}{77}{641}.
\item\label{DirectPhoton1}
   P.~Aurenche et al., \pr{D39}{89}{3275};
   \\
   M.~Bonesini et al., WA70 Collaboration, \zp{C38}{88}{371};
   \\
   G.~Alverson et al., E706 Collaboration, \prl{68}{92}{2584}.
\item\label{DirectPhoton2}
   T.~Akesson et al., AFS Collaboration,
   {\it Soviet J. Nucl. Phys.} {\bf 51}(1990)836;
   \\
   J.~Alitti et al., UA2 Collaboration, Preprint CERN-PPE/92-169 (1992).
\item\label{Jet}
   G.~Arnison et al., UA1 Collaboration, \pl{136B}{84}{294}.
\item\label{HeraWorkshop}
   ``Physics at HERA'', Proceedings of the Workshop, DESY, Hamburg,
    eds. W. Buchm\"uller and G. Ingelman (1991)
\item\label{FLmethod}
   A.M.~Cooper-Sarkar et al., \zp{C39}{88}{281}.
\item\label{KunsztP}
   Z.~Kunszt and E.~Pietarinen, \np{B164}{80}{45}.
\item\label{JPSImethod}
   H.~Jung, G.A.~Schuler and J.~Terron, in
   ``Physics at HERA'', Proceedings of the Workshop, DESY, Hamburg,
   eds. W. Buchm\"uller and G. Ingelman, Vol. II, p. 712 (1991).
\item\label{OpenCmethod}
   R.~van Woudenberg et al., in
   ``Physics at HERA'', Proceedings of the Workshop, DESY Hamburg,
   eds. W. Buchm\"uller and G. Ingelman, Vol. II, p. 739 (1991);
   \newline
   E. Brugnera, {\it ibid.}, Vol. I, p. 557.
\item\label{EllisNason}
   R.K.~Ellis and P.~Nason, \np{B312}{89}{551}.
\item\label{SmithNeerven}
   J.~Smith and W.L.~van Neerven, \np{B374}{92}{36}.
\item\label{FMNR}
   S.~Frixione, M.~Mangano, P.~Nason and G.~Ridolfi, in preparation.
\item\label{MNR}
   M.L. Mangano, P.~Nason and G.~Ridolfi, \np{B373}{92}{295}.
\item\label{MNRFT}
   M. Mangano, P.~Nason and G.~Ridolfi, Pisa preprint IFUP-TH-37-92 (1992).
\item\label{jetdefinitions}
   S.~Ellis, Z.~Kunszt and D.~Soper, \prl{69}{92}{3615};\\
   F.~Aversa et al., \zp{C49}{91}{459}.
\item\label{HMRSB}
   P.N Harriman, A.D. Martin, R.G. Roberts and W.J. Stirling,
   \pr{D42}{90}{798}.
\item\label{Abramowicz}
   H. Abramowicz, K. Charchula and A. Levy, \pl{269B}{91}{458}.
\item \label{Aurenche}
   P. Aurenche, P. Chiappetta, M. Fontannaz, J.P. Guillet and E. Pilon,
   \zp{C56}{92}{589}.
\item\label{Besch} \label{PDFLIB}
   H. Plothow-Besch, CERN preprint CERN-PPE/92-123 (1992).
\item\label{SmithNeervenRiemersma}
   S. Riemersma, J. Smith and W.L. van Neerven, \pl{282B}{92}{171}.
\item\label{Vieira}
   J.H. Da Luz Vieira and J.K. Storrow, \zp{C51}{91}{241}.
\item\label{LAC3failsTOPAZ}
   H. Hayashii, TOPAZ Collaboration, talk presented at the XXVIII Rencontres
   de Moriond, 20-27 March 1993.
\item\label{LAC3failsH1}
   M. Erdmann, H1 Collaboration, talk presented at the XXVIII Rencontres
   de Moriond, 20-27 March 1993.
\item\label{Halzen}
   R.~Fletcher, F.~Halzen, S.~Keller and W.~Smith,
   \pl{266B}{91}{183}.
\end{reflist}
\begin{figcap}
\item\label{FullBornFig}
Differential cross sections for charm production, histogrammed in the
logarithm of $x_g=\exp(-y_{\qq})M_{\qq}/E$,
for $m=1.5$~GeV, HMRS~B structure functions for the proton, and
default values for the scale choices. Only the pure photon component
is included.
\item\label{FullHadrFig}
Contribution of the hadronic component of the photon, compared with the
pure photon component, for two extreme choices of the photon structure
functions.
\item\label{ScaledepFig}
Scale dependence of the cross section (pure photon only).
The solid line corresponds to the choice $\mur=\mu_0$ and
$\muf=2\mu_0$, the dotted line to $\mur=\mu_0/2$
and $\muf=2\mu_0$, the dashed line to $\mur=2\mu_0$
and $\muf=2\mu_0$, and the dot-dashed line to
$\mur=\mu_0$ and $\muf=\mu_0/2$. This last curve is not
shown for the smallest mass cut, because the scale goes outside the
range of validity of the structure function parametrization.
\item\label{MassdepFig}
Charm mass dependence of the cross section (pure photon only).
The solid line corresponds to the choice $m_c=1.5\;$GeV,
the dotted line to $m_c=1.8\;$GeV
and the dashed line to $m_c=1.2\;$GeV.
\item\label{RatioFig}
Ratios of cross sections for different invariant-mass cuts.
{\it (a)} Comparison between the pure photon case and the full
(pure + hadronic) one, for two different photon parton
distributions. {\it (b)} Same as {\it (a)}, for a different
invariant-mass cut. {\it (c)} Mass dependence of the ratio
in the pure photon case, patterns as in fig.~4.
{\it (d)} Scale dependence of the ratio in the pure photon
case, patterns as in fig.~3.
\end{figcap}
\end{document}